# Dynamics of mechanical systems with multiple sliding contacts: new faces of Painlevé's paradox

Péter L. Várkonyi
Budapest University of Technology
Műegyetem rkp. 3
H-1111 Budapest, Hungary
phone: +36 20 251 0668
fax: +36 1 463 1773
email: vpeter@mit.bme.hu

## Funding information

This work has been supported by the National, Research, Development and Innovation Office of Hungary under grant 104501.

# Dynamics of mechanical systems with multiple sliding contacts: new faces of the Painlevé paradox

Péter L. Várkonyi

*Abstract:* we investigate the dynamics of finite degree-of-freedom, planar mechanical systems with multiple sliding, unilateral frictional point contacts. A complete classification of systems with 2 sliding contacts is given. The contact-mode based approach of rigid body mechanics is combined with linear stability analysis using a compliant contact model to determine the feasibility and the stability of every possible contact mode in each class. Special forms of non-stationary contact dynamics including "impact without collision" and "reverse chattering" are also investigated. Many types of solution inconsistency and the indeterminacy are identified and new phenomena related to Painlevé's non-existence and non-uniqueness paradoxes are discovered. Among others, we show that the non-existence paradox is not fully resolvable by considering impulsive contact forces. These results contribute to a growing body of evidence that rigid body mechanics cannot be developed into a complete and self-consistent theory in the presence of contacts and friction.

## 1. Introduction

Impacts and friction between solids were among the first topics extensively studied in the history of modern physics. Impact laws were already proposed by Isaac Newton whereas the first known laws of dry friction were set up long before Newton's time by Leonardo da Vinci. Nevertheless modelling systems with impacts and friction still poses a challenge because the inherent unreliability of simple impact and friction laws [3], [22], [24]; and the complexity of the dynamics induced by them [24], [28]. The present work addresses Painlevé's paradox, an important source of complexity.

The piecewise-smooth nature of dry friction suggests that a unilateral, frictional, rigid point contact is in one of 3 possible modes: sliding, sticking or separation (except for those instants when a rigid component hits another). To identify the instantaneous acceleration of a system in the presence of contacts, all combinations of modes for all contacts (briefly: all contact modes of the system) are routinely tested for consistency [14] [25]. A fundamental shortcoming of the contact mode-based approach was already discovered in the late 19[th] century [12], [20]: the number of consistent solutions may be zero, as well as greater than 1. These observations became known as Painlevé's non-uniqueness and non-existence paradox [1] [2]. For planar systems with a single contact, all possible types of indeterminacy and inconsistency have been identified and listed. There are partial results about solution existence and uniqueness in the case of several contacts [11] [23]. Linear complementarity theory offers a characterization of systems, which always have a unique contact mode regardless of the applied external force [4], nevertheless testing this property is co-NP complete, i.e. infeasible for large problems.

In addition to identifying all forms of Painlevé's paradox, much effort has been devoted to the understanding of their physical origin and to their resolution. The paradoxes are consequences of the assumption of rigidity, and thus a common approach of the analysis is contact regularization, i.e. the modelling of contact compliance [1], [6], [11], [13]. The application of this approach to general systems with a single contact [17] as well as to several specific examples (Painlevé's rod [1], [4], [29], [30]; Painlevé-Klein example [1], [7], [11]; and others [15] [19] [20]) uncovered that the non-existence paradox is resolvable by allowing the possibility of "impact without collision" (IWC), i.e. an impulsive contact force with vanishing pre-impact normal contact velocity. The IWC was also found to be a possible scenario in some cases of the non-uniqueness paradox, but never possible in states free from Painlevé's paradoxes.

Contact regularization also allows one to analyze the dynamic stability of consistent contact modes. The stability analysis of a single contact reveals that consistent contact modes may be dynamically unstable. Nevertheless, the stability analysis does not resolve the non-uniqueness paradox due to the possibility of IWCs and multiple stable contact modes. For single-contact systems, the stability properties of each contact mode have been found to be independent of the characteristics of the underlying compliant contact model. Hence, the question of stability is decidable within rigid body theory. The same analysis [17] also offers a partial resolution of the non-uniqueness paradox by demonstrating that single-contact systems never undergo spontaneous contact mode transitions: a systems never changes from contact mode M1 to M2 as long as M1 remains consistent. The only remaining irresolvable situation is the ambiguity of 'triggered' transitions induced by M1 losing its consistency.

As we have seen, most of our knowledge about Painlevé's paradox comes from the analysis of planar systems with a single contact. Many of these properties are believed to be true for more complex systems, nevertheless this intuition may fail: Dupont et al. [6] showed that a unique consistent contact mode may be unstable in the case of multiple contacts. Hence, in contrast to the single-contact case, unusual contact dynamics must occur in some states, which do not suffer from Painlevé's paradoxes. The present paper reports on some of the author's ongoing efforts to deeper understand other properties and the significance of the Painlevé paradox in general multi-contact systems. The standard methodology of the analysis of friction-induced oscillations [6] [8] [9] [10] [16] is combined with the traditional contact mode based approach of rigid body dynamics. After introducing notations in Sec. 2, a classification of planar systems with two frictional, unilateral, sliding contacts is presented. The classification is based on geometric, and inertial properties, values of the friction coefficient, and properties of the external loads as inputs. A complete list of consistent contact modes is given for each class in Sec 3. The new classification is the refinement of a coarser scheme by Ivanov [11], who investigated the following question: "which systems are guaranteed to have a unique consistent contact mode regardless of the external forces?" Sec. 4 is devoted to the stability analysis of contact modes and of IWCs, using contact regularization. A special type of contact dynamics, called "inverse chattering" is investigated in In Sec. 5. The paper is closed by the discussion of novel phenomena (Sec. 6) and by a Conclusions section. The consistency and stability analysis of contact modes has been published in a preliminary version [27] of this paper.

In addition to the almost complete characterization of the dynamics of 2 sliding contacts, the analysis also allows us to tackle the following questions related to the (partial) resolution of Painlevé's paradox in the context of rigid body theory:

(i)   is the stability of contact modes decidable within rigid body theory?
(ii)  does the consideration of "impact without collision" in addition to traditional contact modes resolve the non-existence paradox?
(iii) are spontaneous contact mode transitions impossible in multi-contact systems?

Our efforts also offer a few small steps towards understanding a fundamental problem of contact mechanics:

(iv)  is it possible to create a complete theory free of non-existence issues without the detailed modeling of deformations?

## 2. Basic notations

We consider an arbitrary finite degree-of-freedom mechanical system with two unilateral, sliding point contacts and nonzero initial sliding velocities. Coulomb friction with constant coefficients of friction $\mu_j$ is assumed at the two contacts. For simplicity, it is assumed that all internal or external constraint forces of the system other than the contact forces at the two sliding contacts can be expressed explicitly and thus the corresponding constraints can be removed from the equations of motion.

Let $\mathbf{q}$ denote the vector of generalized coordinates of the system and the elements of vector $\mathbf{d}=[d_1\ d_2]^T$ denote the gaps between two pairs of interacting contact surfaces. From the Euler-Lagrange equations, one can obtains a system of 2 equations of the form:

$$\ddot{\mathbf{d}} = \mathbf{B}(\mathbf{q})\mathbf{f} - \mathbf{a}(\mathbf{q},\dot{\mathbf{q}}) \qquad (1)$$

where dot means derivation with respect to time; $\mathbf{f}=[f_1\ f_2]^T$ is a vector of Lagrange multipliers representing the magnitudes of the (unknown) normal contact forces at the sliding contacts. The tangential components of the same contact forces are $\mu_i f_i$. Vector $\mathbf{a}=[a_1\ a_2]^T$ contains minus 1 times the (known) second time derivative of the gap $\mathbf{d}$ in the absence of contact forces. $\mathbf{B}=[b_{ij}]$ is a (known) „local, normal mobility matrix" of size 2 by 2, defined as follows: $b_{ij}$ is the second time derivative of the gap $d_i$ in response to a sliding contact force with normal component 1 and tangential component $\mu_j$ at contact $j$. $\mathbf{B}$ is usually non-symmetric due to friction.

We also introduce the following related notations: $\alpha$ for the angle between the vectors $\mathbf{a}$ and $[1\ 0]^T$ such that $0 \leq \alpha \leq 2\pi$; $\mathbf{b}_j$ for the $j^{\text{th}}$ column of $\mathbf{B}$. $\mathbf{b}_j$ represents the response of both contacts to the contact force at contact $j$; $\beta_j$ for the angle between the vectors $\mathbf{b}_j$ and $[1\ 0]^T$; and $\beta_j^*$ for the angle between the vectors $-\mathbf{b}_j$ and $[1\ 0]^T$.

## 3. Consistency of contact modes

The primary goal of the contact mode-based approach is to determine stationary values of $\mathbf{f}$ and $\ddot{\mathbf{d}}$, i.e. the contact forces and the instantaneous acceleration of the system. A closed unilateral point contact with zero normal velocity ($d_i = \dot{d}_i = 0$) and with nonzero tangential sliding velocity in a given direction may remain in sliding mode (S mode) with

$$\ddot{d}_i = 0 \tag{2}$$

$$f_i \geq 0 \tag{3}$$

or it may start free flight (F mode) corresponding to

$$\ddot{d}_i \geq 0 \tag{4}$$

$$f_i = 0 \tag{5}$$

Stick and slip in the opposite direction are ruled out by the prescribed initial tangential velocity.

In a system with two contacts, these scenarios determine 4 possible contact modes represented by the two-letter words SS, FF, SF and FS. To determine the consistency of each contact mode, we combine the equations of motion (1) with the relevant equality constraint (2) or (5) to determine the corresponding value of $f_i$ (sliding) or $\ddot{d}_i$ (free flight). A contact mode is called consistent if the relevant inequality constraints (3) or (4) are satisfied by both contacts. The consistency of each contact mode depends on the elements of $\mathbf{B}$ and $\mathbf{a}$; conditions are summarized in Table 1 (with derivation in Appendix A).

*Table 1* consistency conditions of contact modes

| name | description | consistency conditions |
|---|---|---|
| FF | free flight at both legs | $a_1<0$ & $a_2<0$ |
| SF | sliding on leg 1, free flight at leg 2 | $b_{11}a_1>0$ (positive contact force at leg 1) & $b_{21}/b_{11} \cdot a_1 > a_2$ (other leg lifts up) |
| FS | sliding on leg 2, separation at leg 1 | $b_{22}a_2>0$ (positive contact force at leg 2) & $b_{12}/b_{22} \cdot a_2 > a_1$ (other leg lifts up) |
| SS | sliding on both legs | **a** is in the cone spanned by vectors **b**$_1$ and **b**$_2$ |

The consistency conditions summarized in column 3 of Table 1 are decidable based on the following values:

- the index of the quadrant of $\Re^2$ (real plane) containing **b**$_1$, or equivalently, the index of $\beta_1$ in the ascending ordering of the set $\{\beta_1, \pi/2, \pi, 3\pi/2\}$.
- The index of $\beta_2$ in the ascending ordering of the set $\{\beta_1, \beta_1^*, \beta_2, \pi/2, \pi, 3\pi/2\}$
- the index of $\alpha$ in the ascending ordering of the set $\{\alpha, \beta_1, \beta_2, \pi/2, \pi, 3\pi/2\}$

This observation enables us to classify systems by a three digit code *ijk* where $i \in \{1,2...,4\}$; $j,k \in \{1,2...,6\}$. The first two digits will be used to define a coarse classification (from now: classes), previously described by Ivanov [11]. The last digit will be used as refinement (subclasses). Fig. 1 illustrates directions of the **b**$_1$, **b**$_2$, **a** vectors in each class, and columns 1-5 of Table 2 contain the list of consistent contact modes for each subclass based on the previously determined conditions. The new classification scheme is non-redundant, i.e. there are real mechanical systems in all classes and subclasses. For example, a single planar rigid body with two contacts, and its own weight as external load can belong to any of the subclasses if the positions and directions of contacts as well as the friction coefficients are chosen appropriately.

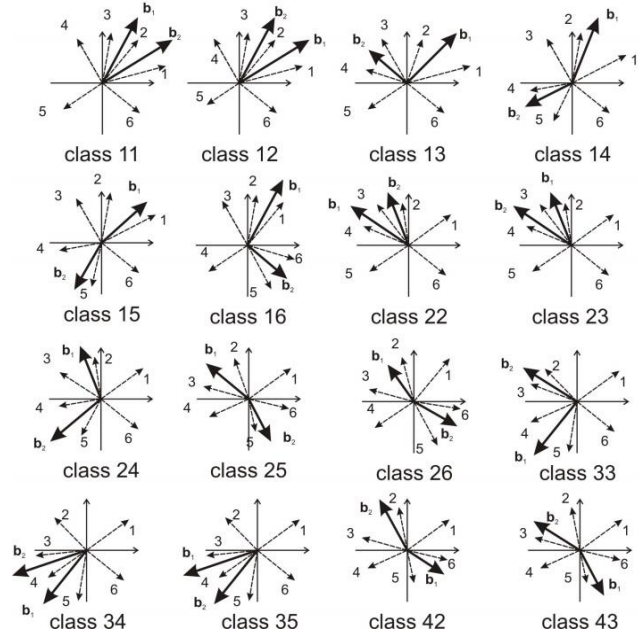

***Fig. 1:*** *directions of $b_j$ (solid arrows) in each of the classes, and the directions of the vector $a$ (enumerated dashed arrows) for every subclass within a class.*

***Table 2*** *Classification of systems based on the directions of vectors $b_1$, $b_2$ and $a$. Column 1 lists classes based on the first two digits of the three-digit labels. A second code in parenthesis indicates the dual of the class obtained by switching the indices of the two contacts. The word 'same' indicates a self-dual class. Columns 2-5 contain lists of subclasses associated with each contact mode. A subclass is in the list if the contact mode is consistent. The stability of the contact mode is marked by dark (unstable) light (model-dependent) and white (stable) background colors. Columns 6-8 show the consistency of all 3 types of IWC in each class by using markers 'x' (consistent), '(x)' (model-dependent) and '-' (inconsistent). Dark background in class 25 indicates that II is consistent but unstable. Columns 9-11 deal with inverse chattering. Marker '?' means that a necessary condition for CC is satisfied, whereas all types of inverse chattering are inconsistent in all other cases*

|           | FF  | SS    | SF    | FS    | IF | FI | II  | CC | CF | FC |
|-----------|-----|-------|-------|-------|----|----|-----|----|----|----|
| **11 (same)** | 5   | 2     | 1 2 6 | 2 3 4 | -  | -  | -   | ?  |    |    |
| **12 (same)** | 5   | 2     | 1 6   | 3 4   | -  | -  | -   | ?  |    |    |
| **13 (41)**   | 5   | 2 3   | 1 6   | 4     | -  | -  | -   | ?  | -  | -  |
| **14 (31)**   | 4 5 | 2 3 4 | 1 6   | 4     | -  | -  | (x) | -  |    |    |
| **15 (32)**   | 4 5 | 1 5 6 | 1 6   | 4     | -  | -  | x   | -  |    |    |

| 16 (21) | 4 | 1 6 | 1 5 6 | 4 5 | - | x | - | - |
|---|---|---|---|---|---|---|---|---|
| 22 (45) | 5 | 3 | 4 5 | 3 4 | x | - | - | - |
| 23 (46) | 5 | 3 | 3 4 5 | 4 | x | - | - | - |
| 24 (36) | 4 5 | 3 4 | 3 4 5 | 4 | x | - | (x) | - |
| 25 (same) | 4 | 3 4 5 | 3 4 | 4 5 | x | x | **x** | - |
| 26 (same) | 4 | 1 2 6 | 3 4 | 4 5 | x | x | - | - |
| 33 (44) | 4 5 | 3 4 | 5 | 3 | - | - | x | - |
| 34 (same) | 3 4 5 | 4 | 5 | 3 | - | - | x | - |
| 35 (same) | 3 4 5 | 4 | 4 5 | 3 4 | - | - | x | - |
| 42 (same) | 4 | 1 2 6 | 5 | 3 | - | - | - | ? |
| 43 (same) | 4 | 3 4 5 | 5 | 3 | - | - | x | ? |

## 4. Contact regularization

In this section we introduce a compliant contact model (Sec. 4.1), which helps us to a more detailed characterization of contact dynamics in each class and subclass. The dynamic stability of the contact modes (Sec. 4.2) is determined with the aid of linear stability analysis. The feasibility of impulsive contact forces, and the robustness of this type of behaviour are also determined (Sec. 4.3).

### 4.1 Contact model

We use a unilateral, Kelvin-Voigt type, linear, viscoelastic contact model with adjustable stiffness in the normal direction, similarly to [6]. The tangential contact force follows Coulomb friction law for sliding. Tangential compliance is not included in the model. The relevance of assuming linear relationship between contact forces and deformations is discussed in Sections 4.2, 4.3.

It is assumed that the mechanical system under investigation behaves as a collection of rigid bodies, which are allowed to overlap in small contact areas with each other. Contact forces emerge if the signed gap $d_i$ between a pair of contact surfaces becomes negative. A contact stiffness scaling factor $\varepsilon$ is introduced such that $\varepsilon \to 0$ corresponds to a rigid contact. The normal contact force is modelled by piecewise linear function of the rescaled contact gap $\bar{d}_i = \varepsilon d_i$ and its derivative $\bar{d}_i'$ with respect to rescaled time $\bar{t} = \varepsilon^{-1/2} t$:

$$f_i = \begin{cases} 0 & \text{if } \bar{d}_i > 0 \\ \max\left\{ \begin{matrix} 0 \\ -k_i \bar{d}_i - q_i \bar{d}_i' \end{matrix} \right\} & \text{if } \bar{d}_i \leq 0 \end{cases} \qquad (6)$$

where the positive scalars $k_i$ and $q_i$ represent rescaled stiffness and damping coefficients of contact $i$.

The usage of the rescaled variables $\bar{d}_i$ and $\bar{d}_i'$ is motivated by the fact that contacts with $O(\varepsilon^{-1})$ stiffness suffer $O(\varepsilon)$ contact deformations under finite forces, and their contact dynamics has an $O(\varepsilon^{1/2})$ characteristic time-scale. The small deformations and the fast characteristic time scale allow the application of singular perturbation techniques. Specifically, the slowly varying parameters $a_i$ and $b_{ij}$ of (1) can be treated as constants during the analysis of fast contact dynamics.

### 4.2 Stability of contact modes

The equation of motion (1) and the compliant contact model (6) together determine the fast dynamics of the contact gaps (**d**) and forces (**f**). Each contact mode corresponds to a stationary value of **f**. Sliding also corresponds to constant values of the corresponding components of **d**. A contact mode being consistent (see Sec. 3) implies that this stationary point exists and satisfies the inequality constraint (3) or (4) for $\varepsilon$ sufficiently small. Nevertheless stationary points of a dynamical system may be dynamically unstable (i.e. repulsive), which means that a system will practically never reach that state. Identifying unstable modes offers a partial resolution of the non-uniqueness paradox.

A contact in free-flight (F) state is special in the sense that the vanishing contact force $f_i=0$ is not sensitive to small perturbations of the system. Thus, on the one hand, the FF mode is always stable (provided that it is consistent). On the other hand, in SF mode, the dynamics of contact 1 can be analysed in itself. Equations (2) and (6) determine a second-order ODE for the time-evolution of $d_1$, which is rewritten as a system of 2 linear, 1$^{st}$ order ODEs:

$$\mathbf{g'} = \begin{bmatrix} 0 & 1 \\ -b_{11}k_1 & -b_{11}q_1 \end{bmatrix} \mathbf{g} - \begin{bmatrix} 0 \\ a_1 \end{bmatrix} \quad \text{where} \quad \mathbf{g} = \begin{bmatrix} \bar{d}_1 \\ \bar{d}_1' \end{bmatrix} \tag{7}$$

As we show in Appendix B, if $b_{11}$ is negative, then the coefficient matrix in (7) has two real eigenvalues, one positive and one negative. The stationary point of the system is a saddle point, i.e. the SF mode is unstable. In the opposite case, the matrix has only negative (real or complex) eigenvalues. Hence the stationary point representing the FS mode is a stable node or a stable focus. Analogously, the FS mode is stable if and only if $b_{22}$ is positive.

For the SS mode, a similar system of ODEs describes the simultaneous dynamics of $d_1$ and $d_2$:

$$\mathbf{g'} = \begin{bmatrix} 0 & 0 & 1 & 0 \\ 0 & 0 & 0 & 1 \\ -\mathbf{BK} & & -\mathbf{BQ} & \end{bmatrix} \mathbf{g} + \begin{bmatrix} 0 \\ 0 \\ \mathbf{a} \end{bmatrix} \tag{8}$$

where

$$\mathbf{g} = \begin{bmatrix} \bar{d}_1 \\ \bar{d}_2 \\ \bar{d}_1' \\ \bar{d}_2' \end{bmatrix} \quad \mathbf{K} = \begin{bmatrix} k_1 & 0 \\ 0 & k_2 \end{bmatrix} \quad \mathbf{Q} = \begin{bmatrix} q_1 & 0 \\ 0 & q_2 \end{bmatrix}.$$

The stability analysis of this type of ODE has attracted considerable attention as a minimal model of break squeal and other types of friction-induced instability [6] [8] [9] [10]. It was found that the stationary point of the dynamics may be destabilized by friction-induced 'mode coupling'. Our systematic analysis of contact mode stability shows the important role of this mechanism.

A stationary point of this system is stable if all eigenvalues of the matrix in (8) have negative real parts. The fulfilment of the stability condition may depend on **B, Q** and **K** (but not on **a** or ε), see Appendix C. There are choices of the angles $\beta_i$, for which the fulfilment of the condition is decidable based on knowing nothing else but $\beta_i$. However we there are other intervals of the angles $\beta_i$ where the stability of the SS mode depends on parameters of the compliant contact model (**K** and **Q**). The results of the stability analysis of all contact modes are summarized in Table 2 and in Fig. 1. Those values of $\beta_i$ for which the SS mode is stable are depicted by dark grey shading in Fig. 1. The curved borders of the stability region in classes 13, 14, 23, 31, 41 and 46 depend on the choice of **K, Q**. The straight boundaries of the stability regions at boundaries of classes are however not sensitive to these parameters, as long as the angles $\beta_i$ are kept constant. Since stability is not affected by **a**, each subclass within a class has the same stability properties.

The stability analysis outlined above does not rely on the ad hoc assumption of the linearity of the force-deformation characteristics (6). The same results can be obtained by using any sufficiently smooth linear or nonlinear contact model of the form $f_i\,(\bar{d}_i, \bar{d}_i')$ provided that it is restoring ( $\partial f_i\,/\partial \bar{d}_i < 0$ ) and dissipative ( $\partial f_i\,/\partial \bar{d}_i' < 0$ ) for all $f_i > 0$ [2] [17]. In the case of a nonlinear contact model, partial derivatives of $f_i$ at the invariant point of the dynamics play the role of parameters $k_i$ and $q_i$ of (6).

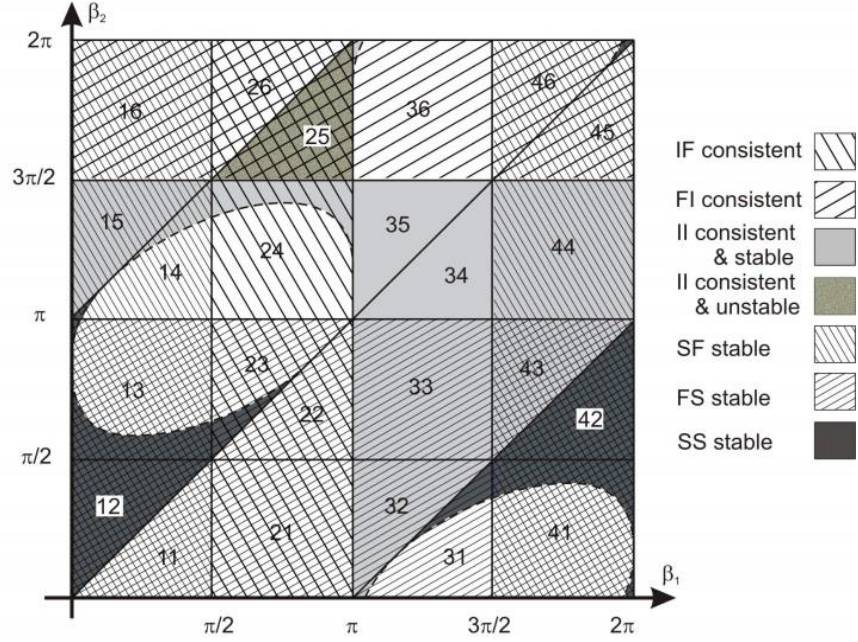

*Fig. 1: Stability regions of the contact modes (SF, FS, SS) and regions of consistency and stability of the three forms of tangential impact (IF, FI, II) in the $\beta_1$-$\beta_2$ plane. The FF mode is always stable (not shown in the figure). The II impact may be stable or unstable, whereas all other IWCs are stable when they are consistent. The dashed boundaries are influenced by $q_1$, $q_2$, $k_1$, $k_2$, $|\boldsymbol{b}_1|$ and $|\boldsymbol{b}_2|$. The actual curves depicted in the figure emerge if $k_1|\boldsymbol{b}_1|/k_2|\boldsymbol{b}_2|=0.5$ and $q_1=q_2=0$. The two-digit numbers refer to labels of the classes described in Sec. 3.*

### 4.3 Impact without collision

In the context of rigid body mechanics, IWCs are modelled as instantaneous impulsive forces, which cause immediate contact stick. Because of the notorious difficulty of modelling simultaneous impacts at multiple contacts within the framework of rigid body theory, we perform an analysis with stiff but compliant contacts. In this context, an IWC is a finite-time process during which $d_i$ quickly diverges towards $-\infty$ for $i=1$ and/or $i=2$ while the corresponding contact force diverges to $+\infty$. The divergence lasts until the tangential velocity of the contact becomes 0, and the contact switches to stick state or slip in the opposite direction. In the current paper, neither the tangential motion of contacts nor the dynamics in stick and reversed slip mode is investigated, because we seek only to detect IWCs instead of predicting their outcomes. Our aim is to find those classes in which an IWC is possible (referred to as consistency of IWC), and those in which an IWC is reached from an open set of initial conditions (stability of IWC). The analysis boils down to investigating the dynamics induced by the linear differential equations (7) and (8). The results obtained below are summarized in rows 6-8 of Table 1.

Three 'modes' of IWC are examined: in mode IF and FI only one support exhibits an increasing contact force, while the other one separates from the ground, whereas in mode II, both legs exhibit impulses. These modes correspond to solution trajectories with the following properties:

- IF: $d_1$ and $d_2$ converge to $-\infty$ and $+\infty$, respectively
- FI: same as IF but with the indices $_1$ and $_2$ swapped
- II: $d_1$ and $d_2$ converge to $-\infty$.

IWCs at a single contact (in our systems: IF and FI modes) have been studied extensively, and the predictions of rigid and of compliant models are identical. Hence we just briefly review these results. IF is possible when the contact force of contact 1 squeezes the contact into the ground ($b_{11}<0$). Because the systems analysed here have a second contact point, it is also necessary that the force at contact 1 lifts up contact 2 ($b_{12}>0$). The IF is stable if the matrix in (7) has a real, positive dominant eigenvalue, which is always true when $b_{11}<0$ (see Appendix B). The consistency and stability condition of FI are analogous.

Simultaneous IWCs at two legs (II) have not been analysed before. Analogously to the single-contact case, an II type IWC occurs when a combination of the two contact forces at the sliding contacts squeezes both contacts into the ground. A trivial necessary condition of the consistency of the II mode is the existence of positive contact forces $f_1$ and $f_2$, which generate negative normal acceleration at both contact surfaces. In other words, the cone spanned by vectors $\mathbf{b}_1$ and $\mathbf{b}_2$ must intersect the negative-negative (third) quadrant of $R^2$. This condition is satisfied by systems belonging to classes 14, 15, 24, 25, 33, 34, 35, 43 and their duals. However, as we show below, this condition is not sufficient: indeed the consistency of II may depend on how contact deformations are modelled.

A closer look at the piecewise linear equation (8) reveals that the II mode is consistent if the matrix of coefficients in (8) has at least one eigenvalue, which is positive (enabling trajectories diverging from 0) and real (such that the sign of $d_i$ does not alternate along the diverging trajectories); furthermore the two coordinates of the corresponding eigenvector have the same sign (enabling trajectories along which $d_1$ and $d_2$ are both negative). The stability of the II type IWC requires that the conditions of consistency are satisfied by the *dominant eigenvalue* of the matrix. Similarly to the SS mode (Sec. 4.1), there are many classes in which the II mode has the same consistency and stability properties for any system (see Fig. 1, Table 2). However in classes 14, 24 and their duals, whether or not the consistency condition is satisfied depends on **B**, **K** and **Q**, see Appendix C. The model-dependent boundaries of consistency are denoted by dashed lines in Fig. 1.

The analysis of IWCs exploits the linearity of $f_i(\bar{d}_i, \bar{d}_i')$. This assumption has no strong physical grounds because real force-displacement characteristics are often strongly nonlinear and they depend on the local shapes of the contact surfaces. Our choice was motivated by its simplicity, and it is justified by the nature of the results derived from the analysis in Sec. 6: the main contribution of the paper is to demonstrate the possibility of special dynamic phenomena (rather than proving that certain things never happen). Phenomena predicted by idealized and simplified contact models, are also likely to be possible in the richer variety of real physical systems.

## 5. Inverse chattering

"Chattering" is a sequence of impacts of decreasing intensity, with a right accumulation point in time (sometimes termed Zeno point). The simplest example of chattering is the bouncing motion of a partially elastic, rigid bouncing ball. Chattering is an important form of contact dynamics and the only mechanism by which sustained contact between rigid bodies is established. Inverse chattering [17] is a similar process, in which the amplitudes of subsequent bounces gradually increase, and thus impacts have a left accumulation point in time. The energy transfer into bouncing mode is typically provided by frictional forces. Inverse chattering is one of the mechanisms by which sustained contact can cease to exist, and it is a source of infinite non-uniqueness of the solution, because at the time when inverse chattering is initiated, the phase of the resulting bouncing motion is unpredictable. We examine here whether or not this type of dynamics is possible in a system with two sliding contacts. In the context of compliant contact models, inverse chattering corresponds to complex, piecewise linear dynamics in a four dimensional state space, because contacts switch between S and F states repeatedly. The apparent difficulty of the analysis motivates us to return to the assumption of rigid contacts. Inverse chattering includes impacts, which is the reason that we review some important general features of common rigid impact models in Sec. 5.1, as well as an approach of dealing with simultaneous impacts. This is followed by developing a necessary condition of consistency in Sec. 5.2. This condition is based on the properties of Sec. 5.1, but it is not specific to a particular impact model. In Sec. 6, we also point out limitations of the necessary condition, emanating from simultaneous impacts. The stability analysis of chattering motion is beyond the scope of this work.

### 5.1 Some features of sliding impacts

In an early stage of inverse chattering motion, the impacts in the inverse chattering sequence are infinitesimally small, while the contacts of the system have a finite initial sliding velocity. Hence, it follows that these impacts are "sliding impacts", i.e. they do not induce contact sticking or reversal of sliding direction. Under these restrictions, almost all simple models of frictional, planar, single-point impact become identical, see for example the model of Chatterjee and Ruina [2]; Kane and Levinson [29]; Pfeiffer and Glocker [23]; Routh (or Poisson) [24]; and Stronge [26]. We will use two important features of these impact models:

(i) a sliding impact at contact point $j$ is not possible unless $b_{jj}>0$.
(ii) the kinematic coefficient of restitution $e$ is always below 1. (In contrast, $e > 1$ is possible in the case of sticking impacts).

In the case of simultaneous contacts at multiple points, the lack of standard impact models [4] motivates us to adopt an ad hoc modelling assumption often used elsewhere: simultaneous impacts are replaced by (finite or infinite) sequences of single point impacts until no pair of contact surfaces involved in the impact moves towards each other ($\dot{d}_i \geq 0$).

### 5.2 Partial consistency analysis

We consider two basic types of inverse chattering. Either all impacts in the sequence occur at one leg while the other one is separated from the ground (CF and FC modes) or the sequence of impacts include repeated impacts at both legs (CC mode).

The case of single-contact inverse chattering has been studied in [17]. They have shown that for the CF and FC modes to be consistent, it is a necessary condition that $e>1$ otherwise the sequence of impacts would decrease exponentially. Hence, the CF and FC modes are inconsistent during sliding motion according to property (ii) of Sec. 5.1.

For the CC mode, it is a necessary but not sufficient condition that the sliding contact force at leg 1 does not squeeze the contact into the ground ($b_{11}>0$), otherwise an impact at leg 1 would cause an IWC with finite impact momentum instead of inverse chattering. The analogous condition $b_{22}>0$ must also be satisfied. The two necessary conditions imply that the CC mode is inconsistent in 18 out of 24 (see column 12 of Table 2). In the remaining 6 classes, a more detailed analysis would be possible with a compliant contact model, which is beyond the scope of the present work. Nevertheless we show some examples of numerical simulation in Sec. 6, where CC-type behaviour is observed.

## 6. Painlevé paradoxes and novel phenomena

The classification presented in previous sections reveals many examples of Painlevé's paradoxes. For example, in class 152, none of the contact modes is consistent. Non-existence is resolved by the fact that one type of IWC is possible in this class. In class 254, all 4 regular contact modes are consistent but only 1 of them is stable. Despite the uniqueness of the stable contact mode, the non-uniqueness paradox is not resolved, because 2 types of IWC are also consistent and stable. Both examples appear qualitatively similar to classical examples of non-uniqueness and non-existence in the literature. At the same time, we also observed several novel phenomena, which are discussed below.

### 6.1 Fast contact dynamics in the absence of the Painlevé paradox

It was demonstrated by [6] that the presence of a unique consistent contact mode does not imply its stability for systems with multiple contacts. This result is confirmed by our systematic analysis, which shows that all systems belonging to subclasses 261, and 262 have this property. Our analysis goes beyond [6] by showing that the IF and FI type IWCs are stable in this case. Hence, such systems most likely undergo an IWC.

### 6.2 Spontaneous contact mode transitions

We have seen that the SS mode has stability boundaries within classes 13, 14, 23, 31, 41 and 46. As the angles $\beta_j$ vary during the motion of a system within a subclass, the system may cross the stability boundary. At the same time, the consistency of a contact mode never changes in the interior of a subclass. This property means that a contact mode may become unstable without losing consistency, and thus multi-contact systems may undergo 'spontaneous' contact mode transitions. The opposite is known to be true for single-contact systems [17] .

### 6.3 Stability analysis of contact modes

The shapes of stability boundaries inside classes 13, 14, 23, 31, 41 and 46 (dashed lines in Fig. 1) depend on several parameters of the compliant contact model. This fact means that the question of stability of the SS mode is often model-dependent, hence undecidable within the framework of rigid body theory. The undecidability of the stability of a contact mode is a new phenomenon, which is not present in systems with a single contact.

### 6.4 A novel form of non-existence resolved by inverse chattering

For all simple examples, which have been investigated in the past, an IWC resolves the non-existence paradox. The new classification summarized in Table 2 is at first sight consistent with this observation. At the same time, there are ranges of parameter values within classes 132,133,142 and 143 (inside the dashed curves of Fig. 1), in which none of the contact modes, nor any type of tangential impact is stable. Hence, such systems must undergo some other form of fast contact dynamics. In these ranges of parameter space, the matrix in (8) has complex eigenvalues with positive real parts. As long as two compliant contacts are active, **d** undergoes linear oscillations of gradually increasing amplitude, until $d_1$ or $d_2$ becomes positive. What follows after this point was examined by numerical simulation of (1)and (8). In the case of a system belonging to subclass 132, we found unboundedly growing oscillations, during which both legs repeatedly hit the ground (Fig. 2a). This motion resembles CC-type inverse chattering, which suggests that inverse chattering is the most likely resolution of solution non-existence in this case. For the sake of comparison, numerical simulations with a set of slightly different parameter values have also been performed. In this case, the SS mode is stable, which is confirmed by the simulation (Fig. 2b). Nevertheless, inverse chattering remains a possible scenario despite the (local) stability of the SS mode (Fig. 2c).

### 6.5 Another form of non-existence not resolved by inverse chattering

An even more puzzling form of solution non-existence appears in subclasses 142 and 143: all types of contact modes, IWCs and inverse chattering are predicted either inconsistent or unstable. Numerical simulations using the compliant contact model have been performed to explore the dynamics of contacts in this special situation. Two types of behaviour have been observed:

(i) gradually growing oscillations with repeated *simultaneous* impacts at the two legs (Fig. 2d).
(ii) sustained microscopic oscillations, during which the legs repeatedly hit the ground and detach. (Fig. 2e).

Scenario (i) is by definition a special form of inverse chattering resembling the CC mode. The CC may not occur in class 14 according to the analysis presented in Sec. 5.2. This contradiction between the theoretical predictions and the numerical results is caused by the fact that our model of simultaneous impact at multiple rigid contacts (Sec. 5.1) does not conform with the $\varepsilon \to 0$ limit of the compliant contact model (6). This example illustrates why the lack of reliable impact models (especially for simultaneous impacts) makes the analysis of fast contact dynamics difficult if not impossible. Scenario (ii) appears in the eyes of a macroscopic observer as sustained sliding accompanied by small-amplitude, high-frequency vibration. The emergence of this novel form of friction-induced oscillation shows that the previously discussed list of possible dynamic behaviours (including classical contact modes, IWCs and inverse chattering) is incomplete without the consideration of other types of contact dynamics.

Our results show the difficulty of developing the "contact mode approach" into a complete system free of the non-existence paradox in the case of multi-contact systems. In the single contact case, adding IWC to contact modes eliminates non-existence. Whether or not the same result can be achieved in the multi-contact case by considering a finite set of dynamic behaviours in addition to contact modes, appears completely unexplored. The lack of such a theory is a fundamental limitations of the rigid body approach in the presence of frictional contacts.

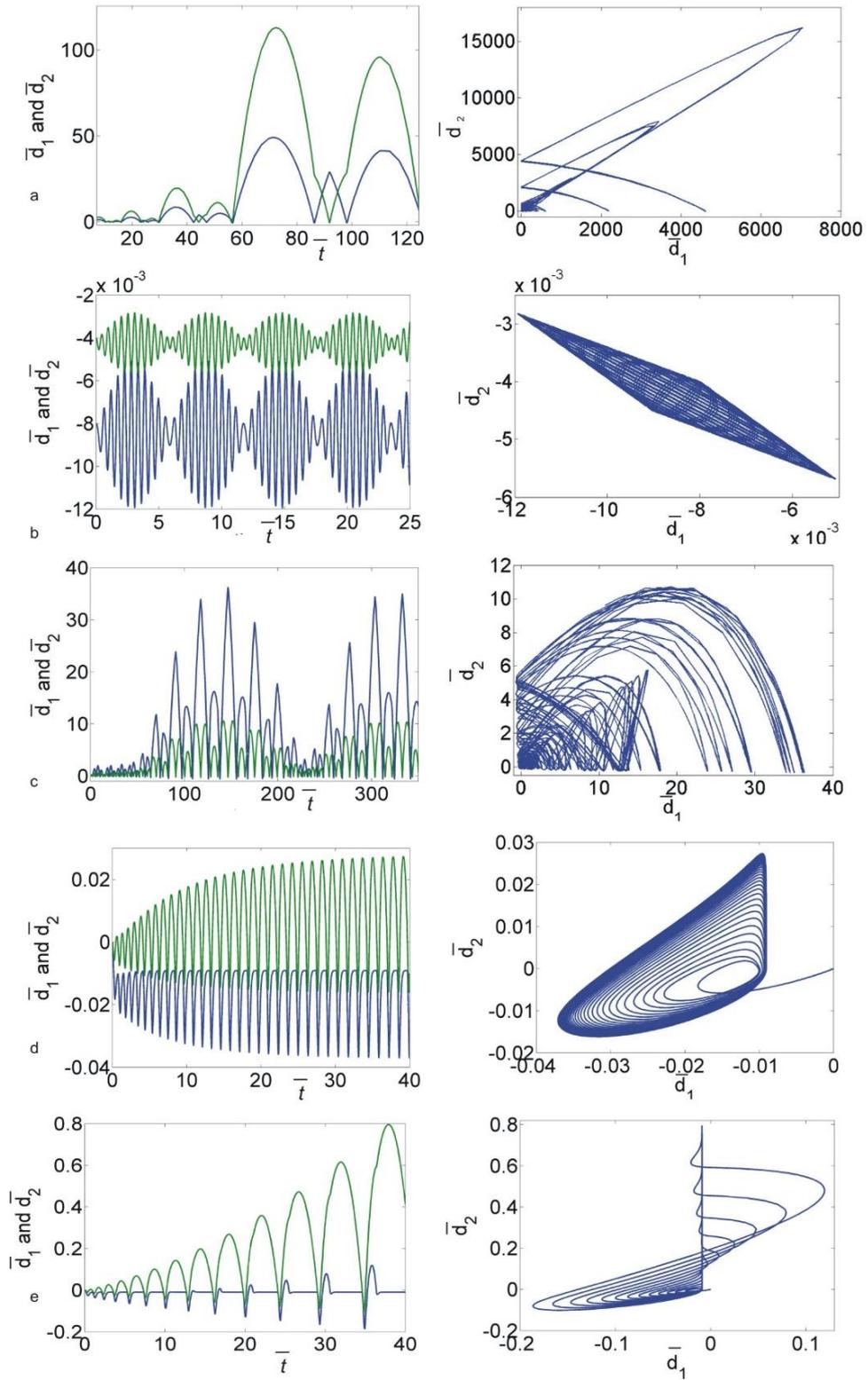

**Fig. 2** numerical simulation of the contact dynamics. a: $|\mathbf{b}_1|=|\mathbf{b}_2|=1$; $\beta_1=0.2\pi$; $\beta_2=0.65\pi$; $k_1=100$; $k_2=200$; $q_1=q_2=0$; $\alpha=0.35\pi$; $|\mathbf{a}|=1$; and initial condition $\mathbf{q}=[-0.008\ -0.004\ \ 0\ \ 0]^T$ (a point belonging to subclass 132). In this case no contact mode or tangential impact is attractive, and the system undergoes inverse chattering-type motion. b: same as a except that $\beta_1=0.05\pi$ (another point of subclass 132). In this case, the SS mode is stable, which is confirmed by the simulation. c: same as b but with initial condition $\mathbf{q}=[-0.05\ -0.05\ \ 0\ \ 0]^T$. Despite the local stability of the SS mode, the system exhibits inverse chattering-type motion. d: $|\mathbf{b}_1|=|\mathbf{b}_2|=1$; $\beta_1=0.15\pi$; $\beta_2=1.07\pi$; $k_1=100$; $k_2=200$; $q_1=q_2=20$; $\alpha=0.2\pi$; $|\mathbf{a}|=1$; and initial condition $\mathbf{q}=[0\ \ 0\ \ 0\ \ 0]^T$. The contacts undergo sustained small amplitude oscillation with repeated impacts e: same as d, except $\beta_2=1.1\pi$. The dynamics resembles inverse chattering, with repeated simultaneous impacts at the two contact points.

## 7. Discussion

In this paper, we attempted to extend classical methods in the analysis of Painlevé's paradox to the case of systems with 2 point contacts. A new classification of such systems has been developed and the consistency and stability of all contact modes and IWCs has been analysed. A partial analysis of inverse chattering has also been performed. Our analysis sheds light on several qualitative differences between single contact systems and multi-contact systems.

The most important results of the analysis was the refutation of the myth that impacts always resolve the non-existence paradox. While this is true in the case of single-contact systems, we have showed the converse in the multi-contact case. We also made an attempt to resolve non-existence by considering inverse chattering in addition to contact modes and impacts. Nevertheless, our efforts remain partially inconclusive due to increasing difficulties of modelling simultaneous impacts. It appears that the extension of the contact mode - based approach to systems with multiple contacts is hopeless even if solution non-uniqueness is acceptable. This finding stresses the limitations of rigid body theory when dealing with contact problems.

Other results also include demonstration of the possibility of spontaneous impacts, and the undecidability of the stability of certain contact modes within rigid body theory. Both findings indicate that techniques offering partial resolution of the non-uniqueness paradox in the single contact case, may fail in the presence of multiple contacts.

The focus of this paper was limited to "instantaneous" phenomena, which are observable within an infinitely short time interval. These include instantaneous accelerations of the system (contact modes), fast contact dynamics (inverse chattering) and singularities of contact dynamics (IWCs). Nevertheless, they do not include singularities of the (much slower) dynamics of sliding motion [4] [19]. Dynamic jamming of multi-contact systems and related phenomena will be the subject of future work.

## Appendix A: consistency conditions of contact modes

**FF mode**: in this mode, (5) implies **f**=0, and $\ddot{\mathbf{d}} = -\mathbf{a}$ by (1). According to the constraint (4), the FF mode is consistent, if $a_1$, $a_2$<0.

**SF mode**: in this mode, (2) implies $f_2$=0; from (1) and (2) with $i$=1, we have $f_1$=$a_1/b_{11}$. The constraint (3) for leg 1 is satisfied if $a_1 b_{11}$>0. Furthermore, $\ddot{d}_2 = -a_2 + b_{21} f_1$, i.e. (4) for leg 2 is satisfied if $-a_2 + a_1 b_{21}/b_{11}$>0. The consistency condition of FS is analogous.

**SS mode**: according to (1) and (2) with $i$=1 and 2, $\mathbf{B}\mathbf{f} = \mathbf{a}$. The constraint (3) is satisfied for both legs if **a** is in the cone spanned by the two column vectors $\mathbf{b}_1$ and $\mathbf{b}_2$ of **B**.

## Appendix B: eigenvalues of the coefficient matrix in (7)

The determinant of the matrix (i.e. the product of the two eigenvalues) is $b_{11}k_1$; whereas the trace of the matrix (i.e. the sum of the eigenvalues) is $-b_{11}q_1$.

Hence, for $b_{11}$>0, the determinant is positive and the trace is negative, implying that both eigenvalues have negative real parts. For $b_{11}$<0, the negative determinant implies two real eigenvalues with opposite signs.

## Appendix C: eigenvalues of the coefficient matrix in (8)

The eigenvalues and the eigenvectors of the matrix

$$\mathbf{M} \stackrel{def}{=} \begin{bmatrix} 0 & 0 & 1 & 0 \\ 0 & 0 & 0 & 1 \\ -\mathbf{BK} & & -\mathbf{BQ} & \end{bmatrix} \tag{9}$$

are investigated here. Specifically, we address the following questions:

1. does any of the eigenvalues have positive real part?
2. are there real, positive eigenvalues such that the two coordinates of the corresponding eigenvector have the same sign?
3. is the dominant eigenvalue real, and the corresponding eigenvector in the positive quadrant?

In the special case of **Q**=0, the block structure of **M** implies that its eigenvalues are $\pm\lambda^{1/2}$ and the eigenvectors are $[\mathbf{x}^T \ \lambda\mathbf{x}^T]^T$ where $\lambda$ and **x** are eigenvalues and eigenvectors of the 2 by 2 matrix -**BK**. Hence, the problem is reduced to the eigenvalue analysis of a 2 by 2 matrix, for which simple, closed-form expressions are available. Below, we provide answers to the three questions without detailed proofs:

1. yes, if and only if **-BK** has a positive eigenvalue, i.e. if either its determinant is negative or if its trace is positive. This is impossible in classes 12 and 42; true in some regions within classes 13, 14, 23, 31, 41, 46; and always true in the rest of the classes. The eigenvalues with non-positive real parts are always purely imaginary.
2. yes in classes 15, 25, 32-35, 43, 44; yes in some regions within classes 14, 24; no otherwise
3. same as for question 2, except for class 25, in which there are two positive real eigenvalues, but only the smaller one has an appropriate eigenvector, implying a negative answer.

In the general case **Q**≠0, the problem is equivalent of a quadratic eigenvalue problem with 2 by 2 matrices. The eigenvalues and eigenvectors can be expressed in closed from, however they appear to have a much more complicated structure than for **Q**=0. Instead of analytical calculation, we performed numerical analysis with systematic variations of **K** and **Q**. The analysis suggests that the answers outlined above for **Q**=0 remain true except that the eigenvalues with non-positive real parts are no more purely imaginary, but typically have negative real parts. This is caused by the damping introduced via **Q.**

## References

[1] Le Xuan Anh (2003) Dynamics of Mechanical Systems with Coulomb Friction (Foundations of Engineering Mechanics). Springer, New York.
[2] Champneys AR, Várkonyi PL, (2016) The Painlevé paradox in contact mechanics, ArXiv e-prints 1601.03545.
[3] Chatterjee A, Ruina A (1998) A new algebraic rigid-body collision law based on impulse space considerations. Journal of Applied Mechanics 65, 939-951.
[4] Cottle RW, Pang JS, Stone RE. (1992) The linear Complementarity Problem. Computer Science and Scientific Computing. Academic Press, Boston.
[5] Génot F, Brogliato, B (1999) New results on Painlevé paradoxes. Eur. J. Mech. A: Solids 18, 653–677
[6] Dupont PE ,Yamajako SP (1997) Stability of frictional contact in constrained rigid body dynamics, IEEE Trans. Robotics and Automation 13, 230-236.
[7] Grigoryan SS (2001) The solution to the Painleve paradox for dry friction. Doklady Physics 46, 499–503
[8] Hervé B, Sinou JJ, Mahé H, Jezequel L (2008) Analysis of squeal noise and mode coupling instabilities including damping and gyroscopic effects. Eur. J. Mech.-A/Solids 27, 141-160.
[9] Hoffmann N, Fischer M, Allgaier R, Gaul L (2002) A minimal model for studying properties of the mode-coupling type instability in friction induced oscillations. Mech. Res. Comm. 29, 197-205.
[10] Hultén J (1993) Brake Squeal-A Self-Exciting Mechanism with Constant Friction. Proc. SAE Truck and Bus Metting, Detroit, MI, USA, 932965.
[11] Ivanov AP (2003) Singularities in the dynamics of systems with non-ideal constraints. J. Appl. Math Mech. 67, 185-192.
[12] Jellett, JH (1872) A Treatise on the Theory of Friction. Hodges, Foster, and Co.
[13] Lecornu L (1905) Sur la loi de Coulomb. Comptes-rendus Acad. Sci. Paris 140, 847-848.

[14] Leine RI, Brogliato B, Nijmeijer H (2002) Periodic motion and bifurcations induced by the Painlevé paradox. Eur. J.Mech.-A/Solids 21, 869-896.

[15] Liu C, Zhao Z, Chen B (2007) The bouncing motion appearing in a robotic system with unilateral constraint. Nonlin. Dyn. 49, 217-232.

[16] Neimark YI, Fufayev NA (1995) The Painlevé paradoxes and the dynamics of a brake shoe. J. Appl. Math. Mech. 59, 343-352.

[17] Nordmark A, Dankowicz H, Champneys AR (2011) Friction-induced reverse chatter in rigid-body mechanisms with impacts. IMA J. Appl. Math 76, 85-119.

[18] Or Y, Rimon E (2008) On the hybrid dynamics of planar mechanisms supported by frictional contacts ii: Stability of two-contact rigid body postures. Proc. IEEE Int. Conf. on Robotics and Automation, Pasadena, CA, USA, 1219–1224.

[19] Or Y, Rimon E, (2012) Investigation of Painlevé's paradox and dynamic jamming during mechanism sliding motion. Nonlin. Dyn. 67, 1647-1668

[20] Painlevé P (1905) Sur les lois du frottement de glissement. Comptes-rendus Acad. Sci. Paris 141, 401–405, 546–552.

[21] Papinniemi A, Lai JC, Zhao J, Loader L (2002) Brake squeal: a literature review. Applied acoustics, 63, 391-400.

[22] Payr M, Glocker C (2005) Oblique frictional impact of a bar: analysis and comparison of different impact laws. Nonlin. Dyn. 41, 361-383.

[23] Pfeiffer F, Glocker C (1996) Multibody dynamics with unilateral contacts. In Wiley Series in Nonlinear Science, Wiley, New York.

[24] Stewart DE (2000) Rigid-body dynamics with friction and impact. SIAM Rev. 42, 3-39.

[25] Van der Schaft AJ, Schumacher JM (1998) Complementarity modeling of hybrid systems. IEEE Tr. Autom. Control 43, 483-490.

[26] Várkonyi, P. L., Gontier, D., Burdick, J. W.: On the Lyapunov stability of quasistatic planar biped robots. Proc. IEEE Intl. Conf. Robotics and Automation 63-70 (2012).

[27] Várkonyi PL (2015) Dynamics of rigid bodies with multiple frictional contacts: new faces of Painleves's paradox In Awrejewicz J, Kazmierczak M, Mrozowski J, Olejnik P (Eds.), Dynamical systems: mechatronics and life sciences, Proc. DSTA 2015, 519-530.

[28] Zhang J, Johansson KH, Lygeros J, Sastry S (2001) Zeno hybrid systems. Intl. J. Robust and Nonlin. Control 11, 435-451.

[29] Zhao Z, Liu C, Chen B, Brogliato B (2015) Asymptotic analysis of Painlevé's paradox. Multibody Syst. Dyn. 35,1-21.

[30] Zhao Z, Liu C, Chen B (2008) The Painlevé paradox studied at a 3D slender rod. Multibody Syst. Dyn. 19, 323-343.